\documentclass[prl,aps,floatfix,twocolumn,showpacs,nofootinbib]{revtex4}

\usepackage[dvipdfm]{graphicx}
\usepackage{bm}
\usepackage{amsmath,amssymb}

\input{colordvi.tex}

\everymath{\displaystyle}
\def\pd{{\rm d}}

\def\reff@jnl#1{{\rm#1\/}}
\def\aj{\reff@jnl{AJ}}                  
\def\araa{\reff@jnl{ARA\&A}}            
\def\apj{\reff@jnl{ApJ}}                
\def\apjl{\reff@jnl{ApJ}}               
\def\apjs{\reff@jnl{ApJS}}              
\def\ao{\reff@jnl{Appl.Optics}}         
\def\apss{\reff@jnl{Ap\&SS}}            
\def\aap{\reff@jnl{A\&A}}               
\def\aapr{\reff@jnl{A\&A~Rev.}}         
\def\aaps{\reff@jnl{A\&AS}}            
\def\azh{\reff@jnl{AZh}}                
\def\baas{\reff@jnl{BAAS}}              
\def\jrasc{\reff@jnl{JRASC}}            
\def\memras{\reff@jnl{MmRAS}}           
\def\mnras{\reff@jnl{MNRAS}}            
\def\newastro{\reff@jnl{New Astron.}}   
\def\pra{\reff@jnl{Phys.Rev.A}}         
\def\prb{\reff@jnl{Phys.Rev.B}}         
\def\prc{\reff@jnl{Phys.Rev.C}}         
\def\prd{\reff@jnl{Phys.Rev.D}}         
\def\prl{\reff@jnl{Phys.Rev.Lett}}      
\def\pasp{\reff@jnl{PASP}}              
\def\pasj{\reff@jnl{PASJ}}              
\def\qjras{\reff@jnl{QJRAS}}            
\def\skytel{\reff@jnl{S\&T}}            
\def\solphys{\reff@jnl{Solar~Phys.}}    
\def\sovast{\reff@jnl{Soviet~Ast.}}     
\def\ssr{\reff@jnl{Space~Sci.Rev.}}     
\def\zap{\reff@jnl{ZAp}}                
\def\nat{\reff@jnl{Nature}}             
\def\physrep{\reff@jnl{Phys.~Rep.}}     
\def\prog{\reff@jnl{PThPS}}        

\begin{document}

\title{Precise Estimation of Cosmological Parameters Using a More Accurate Likelihood Function}

\author{Masanori Sato$^{1,2,}$\footnote{masanori@a.phys.nagoya-u.ac.jp}, Kiyotomo Ichiki$^{1}$, Tsutomu T. Takeuchi$^{3}$}
\affiliation{%
$^{1}$ Department of Physics, Nagoya University, Nagoya 464--8602, Japan
}%
\affiliation{%
$^{2}$ Lawrence Berkeley National Laboratory, 1 Cyclotron Road,
Berkeley, California 94720, USA
}%
\affiliation{%
$^{3}$ Institute for Advanced Research, Nagoya University, Nagoya
464--8601, Japan}%

\date{\today}

\begin{abstract}
The estimation of cosmological parameters from a given data set requires a
construction of a likelihood function which, in general, has a
complicated functional form. We adopt a Gaussian copula and constructed
a copula likelihood function for the convergence power spectrum from a
weak lensing survey. We show that the parameter estimation based on the
Gaussian likelihood erroneously introduces a systematic shift in the
confidence region, in particular for a parameter of the dark energy
equation of state $w$. Thus, the copula likelihood should be used in 
future cosmological observations.
\end{abstract}

\maketitle

{\bf Introduction}.
Currently, a number of wide-field weak lensing (WL) surveys are planned, such
as Subaru Hyper Suprime Cam Survey~\citep{2006SPIE.6269E...9M}, the Panoramic
Survey Telescope \& Rapid Response
System~(Pan-STARRS\footnote{http://pan-starrs.ifa.hawaii.edu/public/}),
the Dark Energy
Survey~(DES\footnote{http://www.darkenergysurvey.org/}), the Large
Synoptic Survey Telescope~(LSST\footnote{http://www.lsst.org/}), the
Joint Dark Energy Mission~(JDEM\footnote{http://jdem.gsfc.nasa.gov/}),
and Euclid~\citep{2010arXiv1001.0061R}, to address questions about the
nature of dark energy and/or the properties of gravity on cosmological
scales.  
To achieve the full potential of these planned future surveys, it is
of great importance to employ adequate statistical measures and methods
of weak lensing for estimating cosmological parameters.

For the cosmological parameter estimation, almost all previous authors
have used the $\chi^2$ method in weak lensing analysis
\citep[e.g.][]{2003ApJ...597...98H,2006ApJ...644...71J,2006A&A...452...51S,2008A&A...479....9F,2010A&A...516A..63S,2010MNRAS.tmp.1452S}.
However, the probability distribution function (PDF) of the weak lensing
power spectrum 
was found to be well approximated by a $\chi^2$ distribution with 
a heavier positive tail than expected from
the normal $\chi^2$ distribution \cite{2009ApJ...701..945S}.  
The $\chi^2$ distribution deviates from Gaussian distribution on large
scales because the number of modes corresponding to the degrees of
freedom are very small.  
Meanwhile, the $\chi^2$ distribution converges Gaussian distribution
at high $\ell$ because of the central limit theorem.  
We have to make maximal use of these pieces of information 
accurately to constrain the cosmological parameters.  
If such information is not taken into account the likelihood function,
the derived cosmological parameters can be systematically biased
\citep{2009A&A...504..689H,2009PhRvD..79b3520I}.

In a companion paper \citep{sit}, we constructed a more properly accurate
likelihood function using the Gaussian copula (hereafter ``copula
likelihood'') for the cosmic shear power spectrum, rather than the
multivariate Gaussian distribution.
We show that the copula likelihood
well reproduces the $n$-dimensional probability distribution of the
cosmic shear power spectrum estimated from 1000 realizations obtained from
ray-tracing simulations performed by \cite{2009ApJ...701..945S}.

In this Letter, we estimate the cosmological parameters
using both the copula and Gaussian likelihoods in order to evaluate 
how the difference between the two likelihoods affects the parameter
estimation.  
Using the Markov Chain Monte Carlo method, we also examine the impact
on both ongoing and future surveys.
 
The cosmological parameters employed for our ray-tracing simulations are
consistent with the WMAP 3-yr results (WMAP3) \citep{2007ApJS..170..377S}.
Detailed descriptions of our ray-tracing simulations are summarized
in \cite{2009ApJ...701..945S} (see also, \cite{2010arXiv1009.2558S}).
 
\vspace{\baselineskip}
{\bf Copulas}.
The likelihood function plays a central role in various statistical analyses.  
In the companion paper \cite{sit}, we derive the copula likelihood
for the cosmic shear power spectrum.
In this section, we briefly summarize the method.

{}From the Sklar's theorem~\citep{sklar1959fonctions}, one can relate any $n$-point
cumulative probability distribution (CDF) to one-point CDFs as
\begin{align}
 &{\rm Prob}(x_{1}\le\hat{x}_{1},x_{2}\le\hat{x}_{2},\dots,x_{n}\le\hat{x}_{n})\equiv
  F(\hat{x}_{1},\hat{x}_{2},\dots,\hat{x}_{n})\nonumber\\
&=C(F_1(\hat{x}_{1}),F_2(\hat{x}_{2}),\dots,
  F_n(\hat{x}_{n})).
\label{sklars_theorem}
\end{align}
Here $\hat{x}_i$ ($i=1,2,\dots,n$) are independent and
identically-distributed observed variables (in this work, the shear power
spectrum, $P_\kappa(\ell)$), $C$ denotes the function called a copula,
$F$ denotes the $n$-point CDF, and $F_{i}$ denotes the one-point CDFs 
(in this work, the cumulative $\chi^2$ distributions).
Thus, the copula describes how one-point CDFs are joined together to 
give an $n$-point CDF. 
A comprehensive proof of Sklar's theorem and rigorous definition of a copula 
are found in \cite{nelsen2006introduction,2010ApJ...708L...9S,2010MNRAS.406.1830T}.

The multivariate Gaussian copula is a copula of $n$-dimensional
random vector that is multivariate normally distributed.
This copula is expressed as 
\begin{equation}
 C(u_1,u_2,\dots,u_n)\equiv\Phi\left(\Phi_1^{-1}(u_1),\Phi_1^{-1}(u_2),\dots,\Phi_1^{-1}(u_n)\right),
\end{equation}
where $\Phi_{1}^{-1}$ is the inverse function of a one-point Gaussian
CDF with mean $\mu_i$ and standard deviation $\sigma_i$, and $u_i\equiv
F_i(\hat{x}_i)$.  
Here $\Phi$ is an $n$-point Gaussian CDF defined by
\begin{align}
&\Phi(\hat{x}_1,\hat{x}_2,\dots,\hat{x}_n)=\int_{-\infty}^{\hat{x}_1}\int_{-\infty}^{\hat{x}_2}\dots\int_{-\infty}^{\hat{x}_n}\frac{1}{\sqrt{(2\pi)^n
  {\rm
  det(Cov)}}}\nonumber\\
&\times\exp\left(-\frac{1}{2}(\boldsymbol{x}-\boldsymbol{\mu})^{\rm T}{\rm
 Cov}^{-1}(\boldsymbol{x}-\boldsymbol{\mu})\right)\pd x_1\pd x_2\dots
  \pd x_n,
\end{align}
with mean $\boldsymbol{\mu}$ and $n\times{n}$ covariance matrix.
${\rm Cov}^{-1}$ shows the inverse covariance matrix.
Hereafter, $\boldsymbol{\mu}\equiv (\mu_1,\mu_2,\dots,\mu_n)$,
$\boldsymbol{x}\equiv (x_1,x_2,\dots,x_n)$, and superscript 'T'
stands for the transpose of vector.

Using this copula, a log-likelihood $\ln\mathcal{L}$ is derived as \cite{sit}
\begin{align}
-2\ln\mathcal{L}_{c}(\hat{x}_1,\hat{x}_2,\dots,\hat{x}_n)&=
\sum_{i=1}^{n}\sum_{j=1}^n(q_i-\mu_i){\rm
 Cov}^{-1}(q_j-\mu_j)\nonumber\\
&-\sum_{i=1}^n\frac{(q_i-\mu_i)^2}{\sigma_i^2}-2\sum_{i=1}^n\ln f_i(\hat{x}_i),
\label{gau_copula_like}
\end{align}
for a general probability distribution, while for Gaussian probability distribution it reduces
\begin{equation}
 -2\ln\mathcal{L}_{g}(\hat{x}_1,\hat{x}_2,\dots,\hat{x}_n)=
\sum_{i=1}^{n}\sum_{j=1}^n(\hat{x}_i-\mu_i){\rm
 Cov}^{-1}(\hat{x}_j-\mu_j) .
\label{gau_like}
\end{equation}
Here, functions $f_i$ represent the derivatives of the CDFs $F_i$,
and we have omitted an irrelevant constant term in the above two equations.
The relation between $u_i$ and $q_i$ is
\begin{equation}
q_i=\sigma_i\Psi_1^{-1}(u_i)+\mu_i,
\end{equation}
where $\Psi_1$ is a cumulative standard normal distribution.
In what follows, we use Eqs.~(\ref{gau_copula_like}) and
Eq.~(\ref{gau_like}) to constrain the cosmological parameters and investigate
how the different likelihood functions cause a difference in the
posterior distribution function of cosmological parameters.

\vspace{\baselineskip}
{\bf Methodology and parameter choices}.
In the case considered in this Letter, the observed variables $\hat{x}_i$
are binned values of the nonlinear convergence power spectrum
$\hat{P}_{\kappa}(\ell_i)$, which are computed using the prescription of
\citep{2003MNRAS.341.1311S} for the nonlinear effects.  
Note that we assume a bin width $\Delta\ln \ell=0.3$, single source
redshift distribution, i.e. all lensed galaxies lie at 
$z_s=1.0$, and we do not consider intrinsic ellipticity dispersion
$\sigma_{\epsilon}$ throughout this Letter.  
Instead, we use the information up to multipole $\ell_{\rm
max}=1071$, because the weak lensing power spectrum estimated from 
a realistic survey on smaller scales than $\ell\sim 1000$ is expected to be
contaminated by the intrinsic ellipticity noise
\citep{2009MNRAS.395.2065T}.  
When constructing the copula likelihood
function, $\mu_i$ and ${\rm Cov}^{-1}$ are estimated from ray-tracing
simulations performed by \citep{2009ApJ...701..945S}.
It is known to be appropriate to adopt a $\chi^2$ distribution for the 
general probability
distribution $f_i$, because the one-point PDF of the convergence power spectrum
is fairly well described by $\chi^2$ distribution with mean and variance of 
$P_{\kappa}(\ell_i)=\langle{\hat{P}_{\kappa}(\ell_i)}\rangle$ and
$\sigma^2(\ell_i)=\langle\hat{P}_{\kappa}(\ell_i)^2\rangle -
P_{\kappa}(\ell_i)^2$, respectively \citep[see,][]{2009ApJ...701..945S}.
{}From \cite{2009ApJ...700..479T}, this $\chi^2$ distribution is written as
\begin{equation}
 f_{\chi^2}(\hat{P}_{\kappa}(\ell_i))=\frac{\hat{P}_{\kappa}(\ell_i)^{\Upsilon -1}}{\Gamma(\Upsilon)}\left(\Upsilon\frac{e^{-\hat{P}_{\kappa}(\ell_i)/P_{\kappa}(\ell_i)}}{P_{\kappa}(\ell_i)}\right)^{\Upsilon},
 \label{eq:chi_square}
\end{equation}
for $\hat{P}_{\kappa}(\ell_i)>0$ and $f_{\chi^2}=0$ for
$\hat{P}_{\kappa}(\ell_i)\le 0$.
Here $\Gamma(x)$ is the gamma function and we define
$\Upsilon\equiv{P_{\kappa}(\ell_i)^2/\sigma^2(\ell_i)}$ which corresponds to
the number of independent modes.

For simplicity, we work with two cosmological parameters in weak
lensing given as 
\begin{equation}
 \boldsymbol{p}=(\Omega_{\rm dm}, \ln 10^{10}\Delta_{\rm R}^2)
\end{equation}
where $\Omega_{\rm dm}$ is the dark matter density today and
$\Delta_{\rm R}^2$ is the amplitude of primordial curvature
perturbations defined at $k=0.002\,{\rm Mpc}^{-1}$
\citep{2007ApJS..170..377S}, whose fiducial values in our analysis are
0.196 and $20.45\times 10^{-10}$, respectively.
The other cosmological parameters are
fixed at the fiducial values of the WMAP 3-yr cosmology.  
When we estimate the equation of state parameter $w$ in the dark energy equation,
combined with another probe of WMAP3 data, we work with
three cosmological parameters given as
\begin{equation}
 \boldsymbol{p}=(\Omega_{\rm dm}, w, \ln 10^{10}\Delta_{\rm R}^2).
\end{equation}
The parameter space we explore is as follows: $\Omega_{\rm
dm}=[0.002,1.0]$, $w=[-6.0,0]$, $\ln 10^{10}\Delta_{\rm R}^2=[-2.5,8.0]$ and
the fiducial values of these parameters are taken as $0.196$, $-1$ and
$3.02$, respectively.
Assuming flat priors for the cosmological parameters,
we employ the Markov Chain Monte Carlo method
\citep{2002PhRvD..66j3511L} to constrain cosmological parameters given
the cosmological observables.  
Sixteen parallel chains were computed and the
convergence test is made based on the Gelman and Rubin statistics called
``$R-1$'' statistics \citep{gelman1992inference}.  
In this work, each chain typically has
300,000 points and $R-1<0.02$ for both of the two models.

\vspace{\baselineskip}
{\bf Results}.
In this section, we constrain the cosmological parameters using
both the copula and Gaussian likelihoods and evaluate how the
difference between the two likelihoods affects the estimation.  
We consider two cases: ongoing and future surveys.

{\it Impact on ongoing weak lensing surveys.}
In our ray-tracing simulations, the survey area is set as
$\Omega_{\rm s}=25$ deg$^2$.
Therefore the fundamental mode of ray-tracing simulations is $\ell_f=72$.

\begin{figure}[!t]
\includegraphics[width=0.45\textwidth]{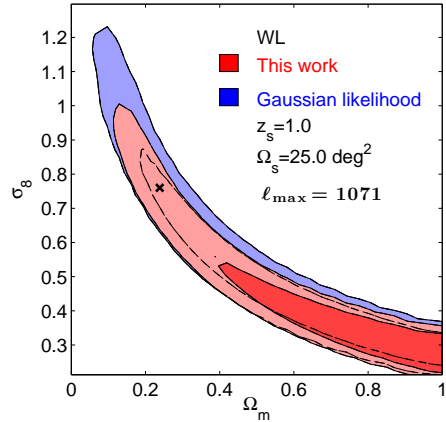}
\vskip-\lastskip
\caption{Two-dimensional constraints (1$\sigma$ and
 2$\sigma$ C.L.) on $\Omega_{\rm m}$ and $\sigma_8$ for survey area
 $\Omega_{\rm s}=25{\rm deg}^2$. The red and blue contours show the
 constraints obtained by likelihood based on the 
 copula likelihood (Eq.~\ref{gau_copula_like}) and Gaussian likelihood
 (Eq.~\ref{gau_like}), respectively.
 The cross symbol shows the fiducial cosmological parameters.
}
\label{fig:om_sigz}
\end{figure}

Figure~\ref{fig:om_sigz} shows 1$\sigma$ and 2$\sigma$
confidence level constraints on the $\Omega_{\rm m}$--$\sigma_8$ plane. 
Note that $\Omega_{\rm m}$ denotes the present-day total matter density, i.e., 
the sum of the current dark matter and baryon densities.  
The red and blue contours 
show the marginalized constraints obtained by the copula likelihood
(Eq.~\ref{gau_copula_like}) and Gaussian likelihood
(Eq.~\ref{gau_like}), respectively.  
Clearly the results from the copula likelihood (red contours) and 
Gaussian likelihood (blue contours) are significantly different.  
The contours obtained with the copula likelihood are shifted toward
a parameter region that gives a lower convergence power compared to
that from the Gaussian likelihood.  
This result is attributed to the fact that
the median of the $\chi^2$ distribution for the convergence power
spectrum is smaller than that of the Gaussian distribution.

\begin{figure}[!t]
\includegraphics[width=0.45\textwidth]{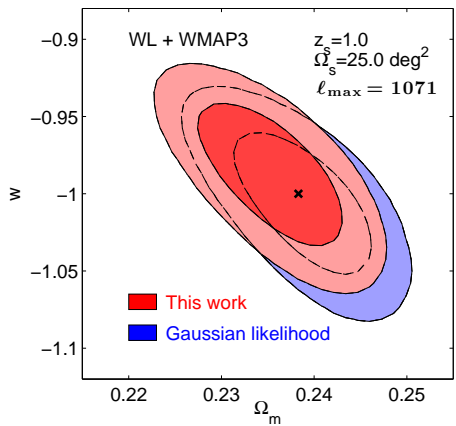}
\vskip-\lastskip
\caption{Same as Fig.~\ref{fig:om_sigz}, but two-dimensional
 marginalized constraints on $\Omega_{\rm m}$ and $w$ for weak lensing plus WMAP
 3-yr results.
}
\label{fig:om_wz20}
\end{figure}

Figure~\ref{fig:om_wz20} shows the constraints on the 
$\Omega_{\rm m}$--$w$ plane obtained when the weak lensing information 
is combined with data from WMAP3. 
Also in this case we see the difference between the results from the two 
likelihoods.
Similarly, the contours obtained by the copula likelihood are shifted 
toward the parameter region that gives the lower power, i.e. smaller $\Omega_{\rm m}$ and
larger $w$, than the Gaussian likelihood, as expected. 
Consequently, if one uses the (approximate) multivariate Gaussian likelihood, 
one will overestimate the lower bound and underestimate the upper
one for the constraint on $w$. 
Specifically, we found an allowed range of the equation of state parameter $w$ 
for the Gaussian likelihood model as
$-1.055 < w < -0.956~,$
while for the copula likelihood model as
$-1.038 < w < -0.941~$
at $95$\% confidence level. 
Therefore, a few percent of systematic error
potentially exists in the parameter estimation if one uses approximate
Gaussian likelihood.

{\it Impact on future weak lensing surveys.}
Next let us assume a survey area of 2000 deg$^2$, which roughly
corresponds to the planned Subaru Weak Lensing Survey \citep{2006SPIE.6269E...9M}. 
Because our ray-tracing simulation is limited to 25 deg$^2$, we have to
extend it to estimate the mean power spectrum $\mu_i$ and their
covariance with help from analytic treatments.
When estimating the mean power spectrum $\mu_i$, we use the linear
perturbation theory based on WMAP 3-yr fiducial cosmology for
multipoles in the range $8 \le \ell < \ell_f=72$, where $\ell_f$
corresponds to the largest angular mode of our ray-tracing simulation
(5$\times$ 5 deg$^2$ for an area).  
For the covariance matrix
${\rm Cov}$, we assume that it has only diagonal elements for multipoles
smaller than $\ell_f=72$ because the linear approximation is valid for that
multipole range.  
In the absence of shape noise, the diagonal elements
would arise from sample variance and should be equal to the power
spectrum squared divided by the number of independent modes in the bin.
Also, we assume that the covariances between the scales larger
than $\ell_f=72$ and scales smaller than $\ell_f=72$ are zero, which means
that the powers at the largest scales are independent from those at
smaller scales.

\begin{figure}[!t]
\includegraphics[width=0.45\textwidth]{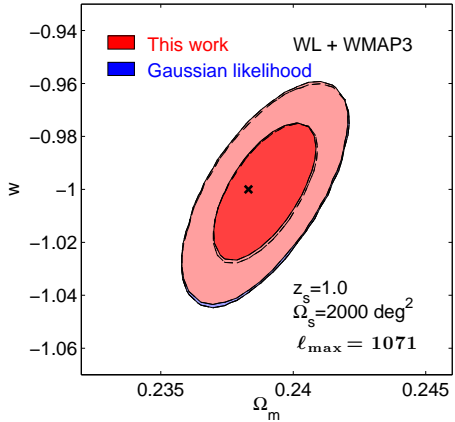}
\vskip-\lastskip
\caption{Same as Fig.~\ref{fig:om_wz20}, but assumed survey area is 2000
 deg$^2$ for weak lensing. This good agreement may not be caused by a physical
reason, but rather attributed to the coarse binning.}
\label{fig:om_wz20_hsc}
\end{figure}

Figure~\ref{fig:om_wz20_hsc} shows the constraints on the ($\Omega_{\rm
m}$, $w$) plane obtained when weak lensing information with the survey
area of $2000$ deg$^2$ is combined with WMAP3.  
In this case, we do not 
see any significant difference between the results from the two likelihoods.  
This result is understood as follows. 
If one considers the larger survey area,
$\Upsilon$ becomes larger at a fixed multipole because $\sigma^2$
becomes smaller (see, Eq.~\ref{eq:chi_square}).  
Hence, $\chi^2$ distribution for the convergence power
spectrum becomes well approximated by the Gaussian distribution at the
fixed multipole. 
Even though at the lowest multipole bin the distribution may
be deviated from Gaussian, the bulk of information comes from the large
multipole bins, where the distribution is almost Gaussian. Hence the
contours become indistinguishable.

It should be noted, however, that the Gaussianity in the distribution of
the power spectrum should be due to our sparse binning of the multipole
space. To utilize the information as much as possible, we have to
divide the multipole space into a larger number of bins than we have
done in this study. In such a case, the distribution of the power
spectrum at each multipole bin will deviate from the Gaussian
distribution and the copula likelihood will still be useful to model the
probability of the power spectrum distribution.

\vspace{\baselineskip}
{\bf Summary}.
In this Letter we constructed a likelihood function for the convergence
power spectrum from weak lensing survey using copula. 
The copula likelihood function describes more accurately
the distribution of the power
spectrum derived from ray-tracing simulations and thus utilizes more
information than the previous methods in the literature. 
We then used
the copula likelihood model to calculate the allowed range of the
cosmological parameters paying particular attention to the dark energy
equation of state parameter $w$, and compare it with that derived from
the conventional multivariate Gaussian likelihood model. 
We found that, 
for the $25$ deg$^2$ weak lensing survey, the results can be different
even combined with the CMB data, depending on which likelihood function
is used. The difference is as large as a few percent.

For the $2000$ deg$^2$ weak lensing survey, we found the results
coincide with each other. This result allows us to use a multivariate
Gaussian likelihood model for the future weak lensing survey, which
will greatly simplify the parameter estimation analysis.  
We note,
however, that this coincidence might be an artifact from our sparse binning
of the multipole space. A full analysis with finer binning is
difficult at present because computational cost is high, and we
leave it for a future study.

\vspace{\baselineskip}
{\bf Acknowledgments.}
We thank Agnieszka Pollo and Masahiro Takada for comments.
We also thank the anonymous referees for careful reading of our
manuscript and very useful and constructive suggestions that help to
clarify our paper further.
M.S. is supported by the JSPS. T.T.T. has been supported
by Program for Improvement of Research Environment for Young Researchers
from Special Coordination Funds for Promoting Science and Technology.
This work is partially supported by the Grant-in-Aid for the Scientific
Research Fund No. 20740105 (T.T.T,), No. 21740177, No. 22012004 (K.I.),
and Grant-in-Aid for
Scientific Research on Priority Areas No. 467 ``Probing the Dark Energy
through an Extremely Wide and Deep Survey with Subaru Telescope''
commissioned by the MEXT of Japan.

\bibliography{ms}

\clearpage

\end{document}